\documentclass[epj]{webofc}
\usepackage[varg]{txfonts}   
\usepackage{hyperref}

\renewcommand*{\eqref}[1]{Eq.~(\ref{eq:#1})}

\newcommand*{\figref}[1]{Fig.~(\ref{fig:#1})}
\newcommand*{\figlab}[1]{\label{fig:#1}}

\newcommand*{\seclab}[1]{\label{sec:#1}}
\wocname{\includegraphics[width=0.25cm,clip]{logoARENA} ARENA2018}
\wocname{ARENA2018}
\woctitle{ARENA2018}
\woctitle{\includegraphics[width=0.25cm,clip]{logoARENA} ARENA2018}
\begin{document}
\title{
	First analysis of inclined air showers detected by Tunka-Rex
	}
%
%

\author{
\firstname{T.} \lastname{Marshalkina}\inst{1},
\firstname{P.A.} \lastname{Bezyazeekov}\inst{1},
\firstname{N.M.} \lastname{Budnev}\inst{1},
\firstname{D.} \lastname{Chernykh}\inst{1},
\firstname{O.} \lastname{Fedorov}\inst{1},
\firstname{O.A.} \lastname{Gress}\inst{1},
\firstname{A.} \lastname{Haungs}\inst{2},
\firstname{R.} \lastname{Hiller}\inst{2}\fnsep\thanks{now at the University of Zürich},
\firstname{T.} \lastname{Huege}\inst{2}\fnsep\thanks{also at Vrije Universiteit Brussel, Brussels, Belgium},
\firstname{Y.} \lastname{Kazarina}\inst{1},
\firstname{M.} \lastname{Kleifges}\inst{3},
\firstname{D.} \lastname{Kostunin}\inst{2},
\firstname{E.E.} \lastname{Korosteleva}\inst{4},
\firstname{L.A.} \lastname{Kuzmichev}\inst{4},
\firstname{V.} \lastname{Lenok}\inst{2},
\firstname{N.} \lastname{Lubsandorzhiev}\inst{4},
\firstname{R.R.} \lastname{Mirgazov}\inst{1},
\firstname{R.} \lastname{Monkhoev}\inst{1},
\firstname{E.} \lastname{Osipova}\inst{4},
\firstname{A.} \lastname{Pakhorukov}\inst{1},
\firstname{L.} \lastname{Pankov}\inst{1},
\firstname{V.V.} \lastname{Prosin}\inst{4},
\firstname{F.G.} \lastname{Schröder}\inst{5, 6},
\firstname{D.} \lastname{Shipilov}\inst{1}
\and
\firstname{A.} \lastname{Zagorodnikov}\inst{1},
(Tunka-Rex Collaboration)
}


\institute{Institute of Applied Physics ISU, Irkutsk, Russia \and
Institut für Kernphysik, Karlsruhe Institute of Technology (KIT), Karlsruhe, Germany \and
Institut für Prozessdatenverarbeitung und Elektronik, Karlsruhe Institute of Technology (KIT), Karlsruhe, Germany \and
Skobeltsyn Institute of Nuclear Physics MSU, Moscow, Russia \and
Institut für Experimentelle Teilchenphysik, Karlsruhe Institute of Technology (KIT), Karlsruhe, Germany \and
Department of Physics and Astronomy, University of Delaware, Newark, DE, USA
          }

\abstract{%
  The Tunka Radio Extension (Tunka-Rex) is a digital antenna array for the detection of radio emission from cosmic-ray air showers in the frequency band of 30 to 80 MHz and for primary energies above 100 PeV.
  The standard analysis of Tunka-Rex includes events with zenith angle of up to 50$^\circ$.
  This cut is determined by the efficiency of the external trigger.
  However, due to the air-shower footprint increasing with zenith angle and due to the more efficient generation of radio emission (the magnetic field in the Tunka valley is almost vertical), there are a number of ultra-high-energy inclined events detected by Tunka-Rex.
  In this work we present a first analysis of a subset of inclined events detected by Tunka-Rex. We estimate the energies of the selected events and test the efficiency of Tunka-Rex antennas for detection of inclined air showers.
}
\maketitle
\section{Introduction}
\label{intro}

The Tunka Radio Extension (Tunka-Rex) is a digital antenna array aimed to detect radio emission from air-showers produced by cosmic rays with energies above 100 PeV in the frequency band of 30-80 MHz \cite{Bezyazeekov:2015rpa}. 
Tunka-Rex requires external trigger and operates jointly with the non-imaging air-Cherenkov light detector Tunka-133 \cite{Prosin:2014dxa} and the  scintillator array of  Tunka-Grande \cite{Budnev:2015cha}. 

Each Tunka-Rex antenna station consists of two perpendicular aligned Short Aperiodic Loaded Loop Antenna (SALLA) \cite{SALLA}, which are designed with decreased sensitivity in lower hemisphere.
This prevents detection of signals reflected from the ground and reduces systematic uncertainties. However, the sensitivity to inclined events is reduced at the same time, which shifts the threshold by about one order of magnitude to the EeV range. 
Moreover, the sensitivity to inclined events is suppressed by the acceptance of the trigger. 
This way we select few high-energy events and present a first analysis of them. 


\section{Event reconstruction}
\seclab{sec-1}
Tunka-Rex is triggered either by Tunka-133 (at clear, moonless nights) or by Tunka-Grande (the rest of the time). In this work we take only events triggered by Tunka-Grande, since Tunka-133 has limited field of view by design (up to zenith angles of $50^\circ$).
The reconstruction of inclined events is similar to the standard one, described in Ref.~\cite{Bezyazeekov:2015ica}, except for few modifications. 
First of all, the signal window is extended from 200 to 500~ns due to geometry reasons: the dimensions of an inclined event are much larger than of a vertical one. 
Secondly, all air-shower parameters, including energy and direction are reconstructed by Tunka-Rex standalone, i.e. Tunka-Grande reconstruction is not included in this analysis.

We analyse events taken from 2015-2017 (424 runs) and set a quality cut on the number of antenna stations with signal. We obtained a subset of 52 events in total for zenith angles from $60^\circ$ to $90^\circ$. 
There are three further cuts applied: 
1) zenith angle $60^\circ$--$70^\circ$ with minimal number of 21 antenna stations (28 events); 
2) $70^\circ$--$80^\circ$ with 16 antenna stations (19 events);
3) $80^\circ$--$90^\circ$ with 10 antenna stations (5 events).

The distribution of the events on the sky can be seen in \figref{fig-1}. Low sensitivity to very inclined events is in agreement with the acceptance of the antenna array and the trigger. 
Moreover, we see a suppression of the flux from North, which can be explained by shadow from mountains with altitude of 1.5–2.5 km in 5-10 km towards North (the asymmetry caused by geomagnetic suppression is not significant for the inclined events and has opposite sign).

\begin{figure}[h]
\centering
\includegraphics[width=0.28\textwidth]{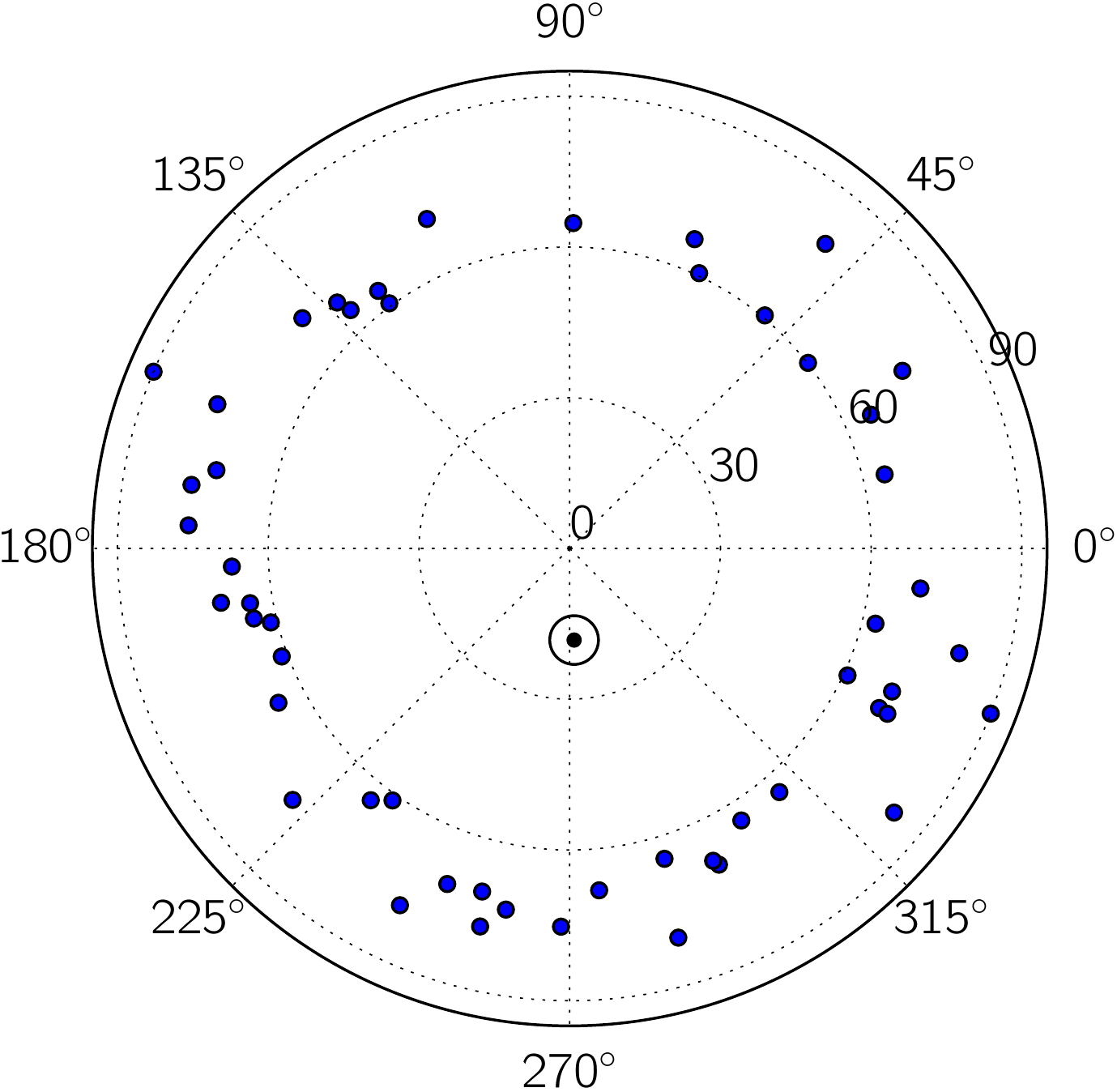}~~\includegraphics[width=0.31\textwidth]{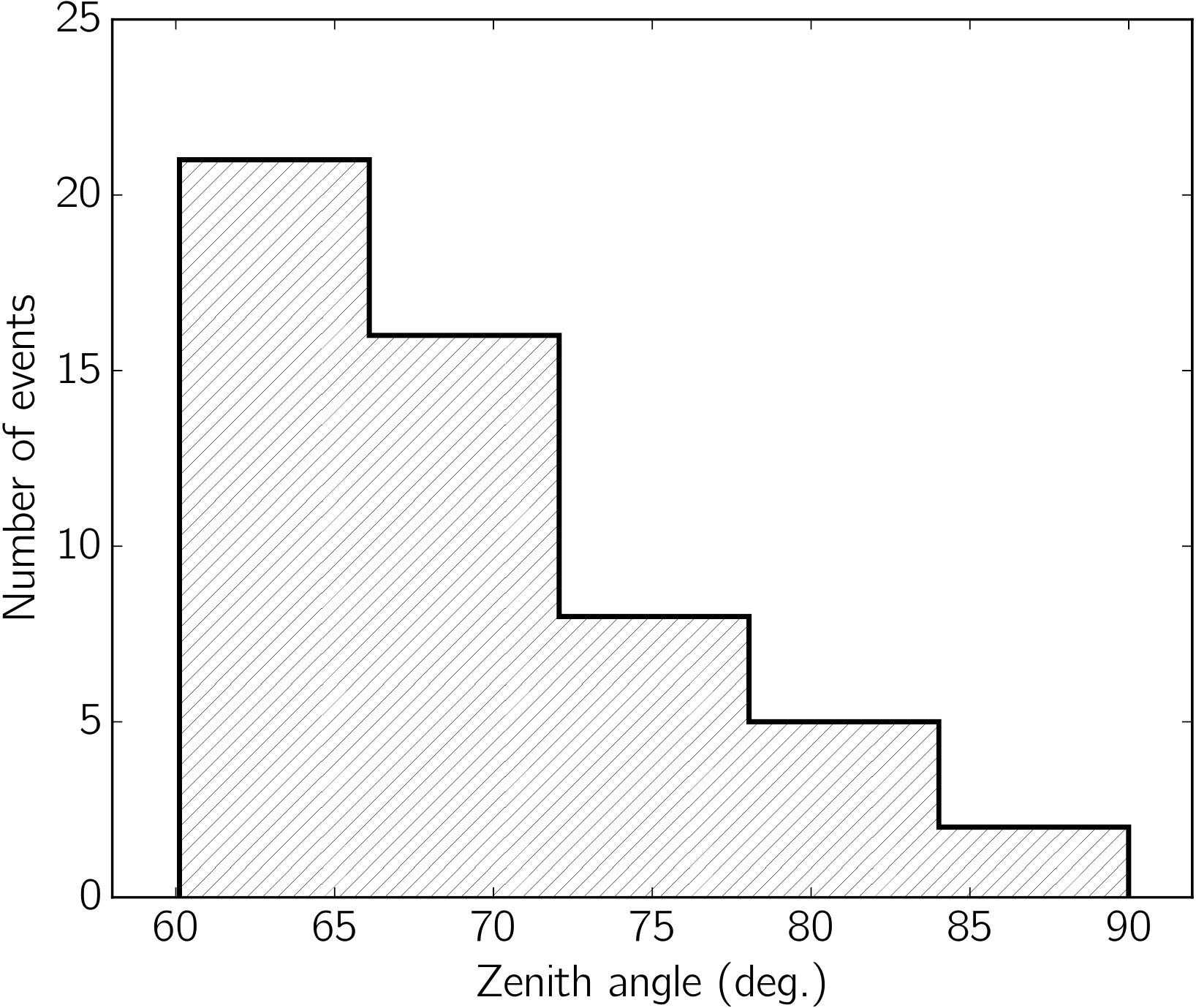}~~\includegraphics[width=0.31\textwidth]{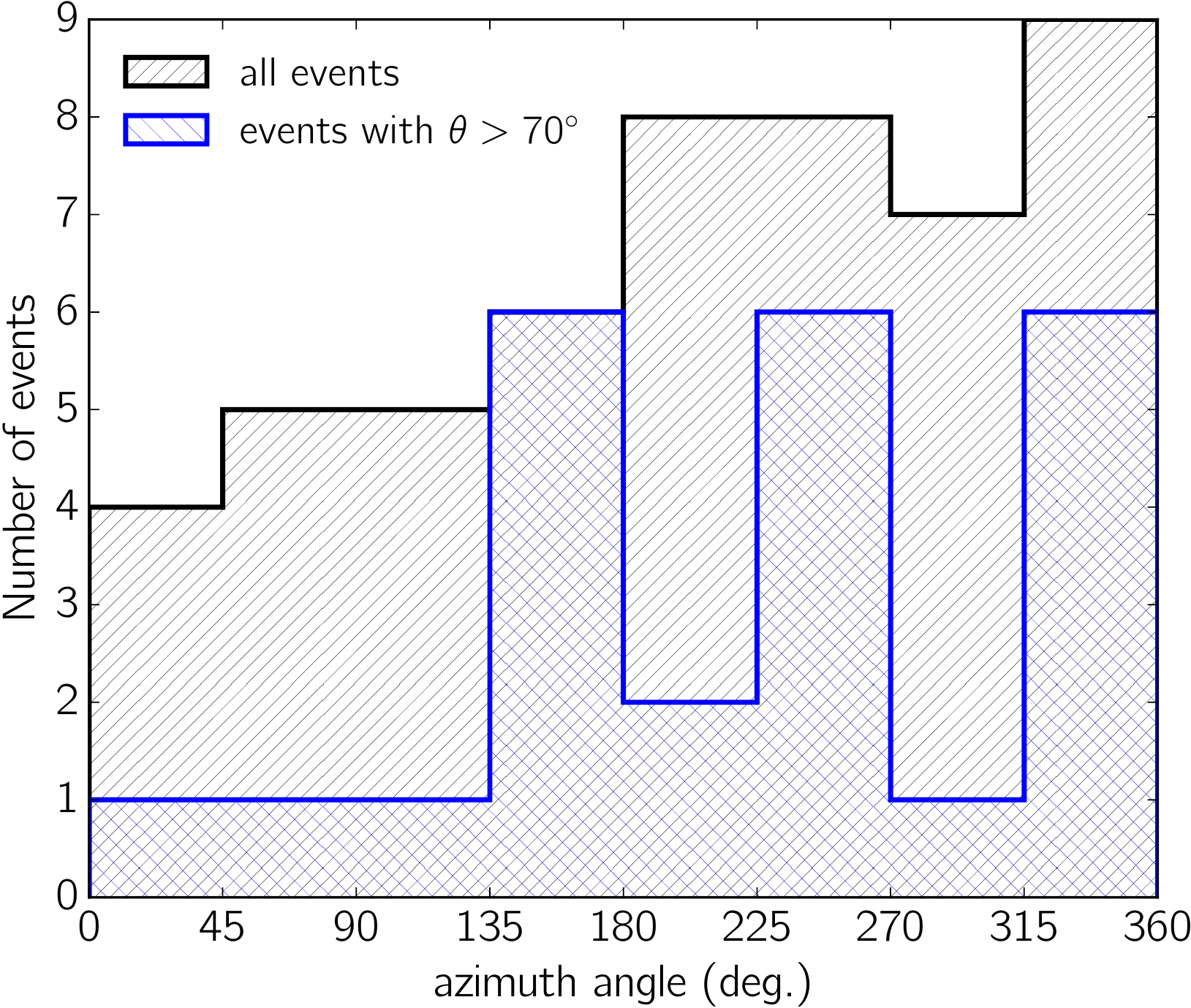}
\caption{\textit{Left}: Distribution of detected inclined events on the sky.  \textit{Center}: Distribution of inclined events as a function of zenith angle. \textit{Right}: Distribution of inclined events as a function of azimuth angle.}
\figlab{fig-1}     
\vspace{-0.4cm}
\end{figure}

\begin{figure}[h]
	\centering
	\includegraphics[width=0.56\linewidth]{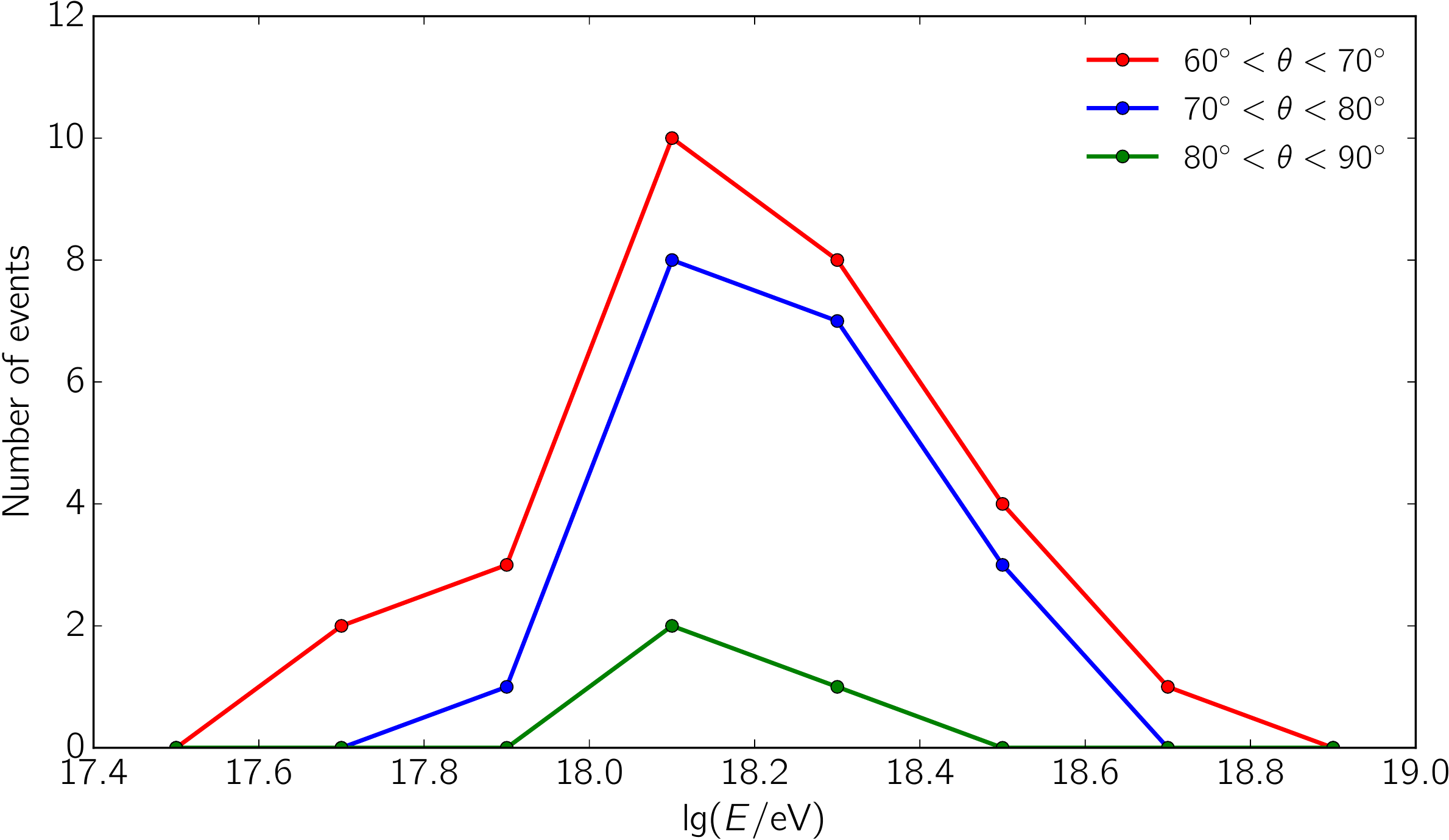}~~~~~
	\includegraphics[width=0.32\linewidth]{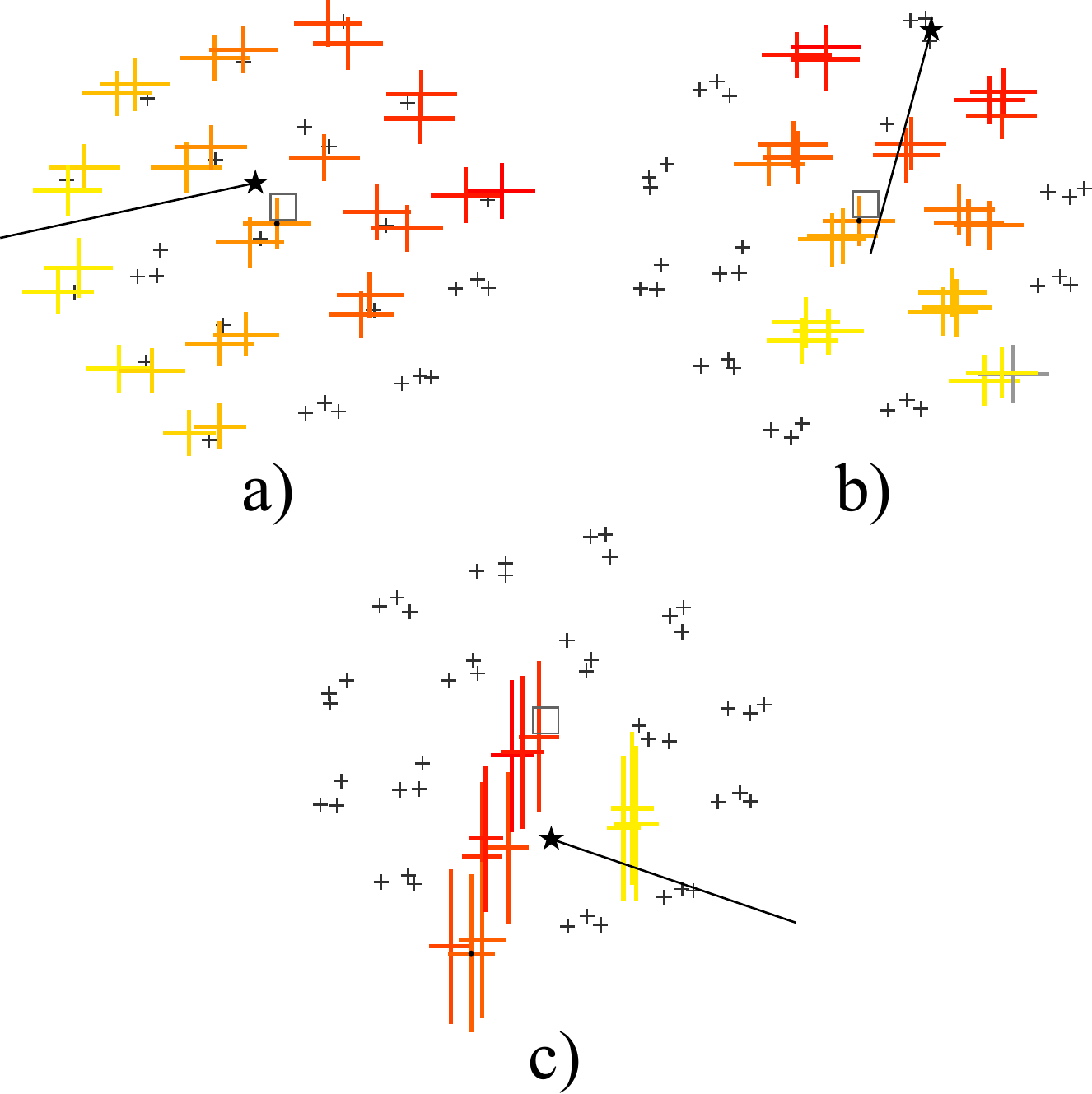}
	
	\caption{\textit{Left}: Rough energy estimation for inclined events.
	\textit{Right}: Examples of inclined events in different zenith angle ranges: a) 
	$60^\circ-70^\circ$, b) $70^\circ-80^\circ$, c) $80^\circ-90^\circ$. 
	Arrows denote the direction of the air-shower.
	Color code of crosses corresponds to the arrival time and their size indicates the signal strength (small ones are the antenna stations which were not read out).}
	\figlab{fig-2}       
	\vspace{-0.4cm}
\end{figure}
We estimate the energy $E$ using the following approximation obtained from the standard Tunka-Rex parametrization~\cite{Kostunin:2015taa}:

\begin{equation}
E=\kappa\times(D_{70^\circ}/D_{50^\circ})\langle \mathcal{E} \rangle,
\end{equation}
where $\kappa=868$~EeV/($\mu$V/m) is the normalization factor for vertical events, $D_{70^\circ}/D_{50^\circ}\approx 4$, $D_{70^\circ}$ is the average distance to shower maximum for an inclination of $70^\circ$ (i.e. inclined events), $D_{50^\circ}$ is the average distance to shower maximum for inclination of $50^\circ$ (i.e. vertical events), $\langle \mathcal{E} \rangle$ is the mean value of detected amplitudes. 
We use $\langle \mathcal{E} \rangle$ as estimator due to a flattening of the lateral distribution function (LDF) and smearing its features (with expected uncertainty of about 20\%--50\%).

The distribution of reconstructed energies is give in \figref{fig-2}. As expected, the threshold is shifted to about $10^{18}$ eV. 
However, precise reconstruction with the LDF method \cite{Kostunin:2015taa} is difficult due to the smeared LDF.
Therefore, other methods of air shower reconstruction are required, for example, a template fit method described in Ref.~\cite{Bezyazeekov:2018yjw}.

\section{Summary}
We present first analysis of inclined air-shower events detected by Tunka-Rex, where one can see few samples in \figref{fig-2}, right. 
Despite of the small number of those events we confirmed that Tunka-Rex equipped with SALLA is sensitive for this type of events. 
However, external circumstances can affect the number of detected inclined events, e.g., the shadow of nearby mountains or the not yet investigated influence of the ground, which has to be studied in more detail. 
Due to smeared LDF, standard methods of reconstruction are not efficient and have low performance. 
Therefore, more sophisticated methods are required for the reconstruction of very inclined events.






\section*{Acknowledgements} 
This work is supported by the Deutsche Forschungsgemeinschaft (DFG Grant No. SCHR 1480/1-1),
Helmholtz grant HRSF-0027, the Russian Federation Ministry of Education and Science (Tunka shared core facilities, unique identifier RFMEFI59317X0005,  agreements: 3.9678.2017/8.9, 3.904.2017/4.6, 3.6787.2017/7.8, 3.6790.2017/7.8), the Russian Foundation for Basic Research (grants 16-02-00738, 17-02-00905, 18-32-00460).

\end{document}